\documentstyle[prl,aps,epsf,twocolumn]{revtex}
\large
\begin{document}
\draft
\twocolumn[\hsize\textwidth\columnwidth\hsize\csname
@twocolumnfalse\endcsname

\title{Phase diagram of disordered spin-Peierls systems}
\author{M. Mostovoy\cite{Perm1},
D. Khomskii\cite{Perm2}, and J. Knoester}
\address{Theoretical Physics and Materials Science Center,\\
University of Groningen, Nijenborgh 4, 9747 AG
Groningen, The Netherlands}
\date{\today}
\maketitle
\begin{abstract}
\widetext
\leftskip 54.8pt
\rightskip 54.8pt

We study the competition between the spin-Peierls and the
antiferromagnetic ordering in disordered quasi-one-dimensional
spin systems.  We obtain the temperature vs disorder-strength
phase diagram, which qualitatively agrees with recent
experiments on doped CuGeO$_3$.

\leftskip 54.8pt
\end{abstract}

]

\narrowtext

The discovery of the first inorganic spin-Peierls (SP) material
CuGeO$_3$ opened the possibility to study the influence of doping
on the SP transition \cite{Hase93.1}. The recently obtained phase
diagram of doped CuGeO$_3$ has several surprising
features \cite{Hase93.2,Renard,Regnault,Hir}. It turns out that
doping, while suppressing the SP state, at the same time induces
long-range antiferromagnetic (AF) order, with the Neel
temperature initially increasing with the doping concentration.
Furthermore, a doping range is found where SP and AF order
coexist.

At first glance, it seems very strange that disorder (doping) may
lead to the enhancement of some order parameter (in this case the
AF one).  Also, the coexistence of the dimerized SP state, in
which spins are bound into singlets, with a spontaneous
subblattice magnetization that requires the presence of free
spins, is rather puzzling.  In this Letter we address both these
issues and obtain a phase diagram that is very similar to the
experimental one.  We also predict a new phenomenon: a
re-entrance transition from the dimerized SP state back into the
undimerized state with decreasing temperature.

Theoretically, the possibility of long-range magnetic order in
doped SP systems was discussed in Refs.\CITE{Fukuyama,KGM,Zeit},
where the lattice was treated classically and it was assumed that
impurities ``cut'' the spin chains into finite segments.  It was
argued that the lattice relaxation in these segments results in
the appearance of regions with a suppressed dimerization (close
to impurities in the model of Ref.~\CITE{Fukuyama}, or centered
at kinks in the lattice dimerization in Refs.~\CITE{KGM,Zeit}).
The AF correlations that develop in these regions may, in
principle, stabilize an inhomogeneous state in which the SP and
AF orders coexist.  The enhancement of the magnetic
susceptibility by disorder-induced kinks was also discussed in
Ref.~\CITE{Fabrizio}.  Although these considerations provide a
qualitative understanding of the magnetic ordering in doped SP
materials, the description of the thermodynamics of the mixed
SP$+$AF state within the same approach is complicated and so far
has not been given.

In this Letter we consider a model that does allow for a detailed
study of the competition between the SP and AF phases in the
presence of disorder.  Instead of considering disorder that
randomly cuts chains into finite segments, we assume that doping
results in small fluctuations of the spin-exchange constants on many
bonds.  Furthermore, we treat the lattice and the interchain
interactions in the ``chain mean field'' approximation
\cite{IPS}.  Then, the effective single-chain Hamiltonian reads:
\begin{equation}
H_{s} = \sum_n
J_{n,n+1} {\bf S}_n\!\!\cdot\!{\bf S}_{n+1}
- h \sum_n (-)^n S_n^z ,
\label{hchain}
\end{equation}
where the exchange constants have the form:
\begin{equation}
J_{n,n+1} = J_0 + (-)^n \Delta + \delta J_{n,n+1} .
\end{equation}
Here, $\Delta$ is the average value of the SP dimerization, and
$\delta J_{n,n+1}$ is the random contribution due to doping.  The
antiferromagnetic order parameter $h$ in Eq.(\ref{hchain}) is the
amplitude of the alternating magnetic field created by the
neighboring chains.  The two order parameters $\Delta$ and $h$
have to be found in a self-consistent way: the dimerization
amplitude $\Delta$ is proportional to the alternating part of
$\langle\langle {\bf S}_n\!\!\cdot\!  {\bf S}_{n+1}
\rangle\rangle$, while $h$ is proportional to the alternating
part of $\langle\langle S_n^z \rangle\rangle$, where
$\langle\langle \ldots \rangle\rangle$ denotes the thermal and
disorder average.

Unfortunately, analytical expressions for these disorder averages
within the model Eq.(\ref{hchain}) are not known and their
numerical calculation at present cannot be done efficiently
enough to solve the self-consistency equations.  This motivates
us to simplify the model, by considering only the XY part of the
spin-spin interaction, {\em i.e.}, we substitute ${\bf
S}_n\!\!\cdot\!{\bf S}_{n+1}$ by $S_n^x S_{n+1}^x + S_n^y
S_{n+1}^y$.  In the absence of disorder the XY model is known to
provide a reasonable description of both the AF and the SP
state \cite{Bul,BBK}. Moreover, as follows from numerical and
analytical studies, in the presence of disorder the
low-temperature behavior of the chain susceptibility to uniform
and alternating magnetic fields is universal, {\em i.e.},
independent of the anisotropy of the spin
exchange \cite{Dasgupta,DFisher,Hyman}.

By means of the Jordan-Wigner transformation the XY model can be
mapped on a free-fermion Hamiltonian.  Then, $J_{n,n+1}/2$
becomes the fermion hopping amplitude between sites $n$ and
$n+1$, while $h$ becomes the amplitude of an alternating on-site
potential.  The coexistence of SP and AF order in the XY
model corresponds to the coexistence of a Peierls dimerization
and an on-site charge density wave in the spinless fermion
model.

In the weak-coupling and weak-disorder limit, {\em i.e.}, for
$\Delta,h,\delta J \ll J_0$, we now introduce a continuum
description of the chain (cf.  Ref.~\CITE{TLM}).  The Hamiltonian
then becomes:
\begin{equation}
H_{f} = \int\!\!dx \psi^{\dagger}(x)
\left[
\sigma_3 \frac{v_F}{i} \frac{d}{dx} +
\sigma_1 (\Delta + \eta(x)) + \sigma_2 h
\right]
\psi(x) ,
\label{H_f}
\end{equation}
where $\sigma_a$ ($a =1,2,3$) are the Pauli matrices.  The first
term in the Hamiltonian $H_{f}$ describes the free motion of
the fermions with the Fermi velocity $v_F = a J_0$, while the
second and the third terms describe the backward scattering
caused by the dimerization, the disorder, and the staggered
magnetic field.  The disorder $\eta(x)$ is related to the
disorder in the spin-exchange constants by
\begin{equation}
\eta(2na) = \frac{1}{2}
\left(\delta J_{2n-1,2n} - \delta J_{2n,2n+1}\right) ,
\label{etaJ}
\end{equation}
where $a$ denotes the lattice constant in the chain direction.
We will assume white noise disorder with a correlator:
\begin{equation}
\label{Gauss}
\langle \eta (x) \eta(y) \rangle = A \delta(x - y) ,
\end{equation}
which corresponds to the statistical independence of the
variations of the exchange couplings on different bonds in the
discrete model Eq.(\ref{hchain}).

In the absence of a magnetic field ($h=0$), the disorder-averaged
density of single-fermion states $\rho(\varepsilon)$ of the
Hamiltonian Eq.(\ref{H_f}) was found analytically in
Ref.\CITE{OE}.  The density $\rho(\varepsilon)$ is a symmetric
function of the energy $\varepsilon$.  Its form depends crucially
on the parameter $g = A / (v_F \Delta)$.  For $g < 2$ the density
of states has a pseudogap (a Peierls gap filled by
disorder-induced states), while for $g > 2$ (strong disorder) the
pseudogap disappears and $\rho(\varepsilon)$ diverges at
$\varepsilon = 0$ ($\rho(\varepsilon) \propto
|\varepsilon|^{\frac{2}{g}-1}$ at $|\varepsilon| \ll \Delta$).

A nonzero alternating magnetic field mixes the $h=0$ eigenstates
with opposite energies and transforms the pair of eigenstates
with energies $\pm \varepsilon$ into a pair of eigenstates with
energies $\pm \sqrt{\varepsilon^2 + h^2}$.  Therefore, the
disorder-averaged $\Omega$-potential ($\Omega_{f} = - T
\langle \ln \Xi_{f} \rangle$, $\Xi_{f}$ being the partition
function of the grand-canonical ensemble of fermions with zero
chemical potential) is given by
\begin{equation}
\Omega_f = -\frac{2}{\beta}
\int_0^W\!\!d\varepsilon
\rho(\varepsilon) \ln \left[ 2 \cosh \left(
\frac{\beta\sqrt{\varepsilon^2 + h^2}}{2} \right) \right] ,
\end{equation}
where $W$ is the energy cut-off.  The two order parameters,
$\Delta$ and $h$, satisfy the self-consistency equations:
\begin{eqnarray}
\Delta &=& - \lambda_{\Delta} \langle\langle \sigma_1
\rangle\rangle = - \lambda_{\Delta} \frac{\partial \Omega_{f}}
{\partial \Delta} , \label{scd}
\\ \nonumber \\
h &=& - \lambda_{h} \langle\langle \sigma_2 \rangle\rangle =
- \lambda_{h} \frac{\partial \Omega_{f}}
{\partial h} ,
\label{sch}
\end{eqnarray}
where $\lambda_{\Delta}$ and $\lambda_h$ are the corresponding
coupling constants describing the interchain interactions.

In the absence of disorder, $\Omega_{f}$ depends on $\Delta$
and $h$ only through the combination $\sqrt{\Delta^2 + h^2}$.  As
a result, the two self-consistency equations acquire the same
(BCS) form; as, however, they have different coupling constants,
they cannot be satisfied simultaneously, unless $\lambda_{\Delta}
= \lambda_h$.  Thus, in agreement with previous
studies \cite{Inagaki}, we find that in the absence of disorder
the AF and SP phases cannot coexist and the phase with the larger
coupling constant is realized.  A competition between these two
phases always exists in spin chain materials and some special
conditions, such as a strong spin-phonon coupling\cite{BBK} or a
significant next-nearest-neighbor interaction \cite{Castilla}, are
necessary for the SP state to win.  This explains why the number
of SP materials is small.

Disorder in the spin-exchange constants suppresses the dimerized
state by filling the SP gap with single-fermion states and thus
reducing the energy gain due to dimerization.  At the same time,
these disorder-induced states enhance the antiferromagnetic
susceptibility of the chains: The effect of an alternating
magnetic field is strongest for the fermionic states with
$|\varepsilon| \leq h$, as the occupied state with energy
$-|\varepsilon|$ is pushed down to $-\sqrt{\varepsilon^2+h^2}$.
The higher is the density of states near $\varepsilon = 0$, the
more energy is gained when AF order appears.  Within the mean
field approximation, this enhancement of the chain magnetic
susceptibility due to disorder results in an increase of the Neel
temperature.

\begin{figure}[htbp]
\centering \leavevmode
\epsfxsize=8cm \epsfbox{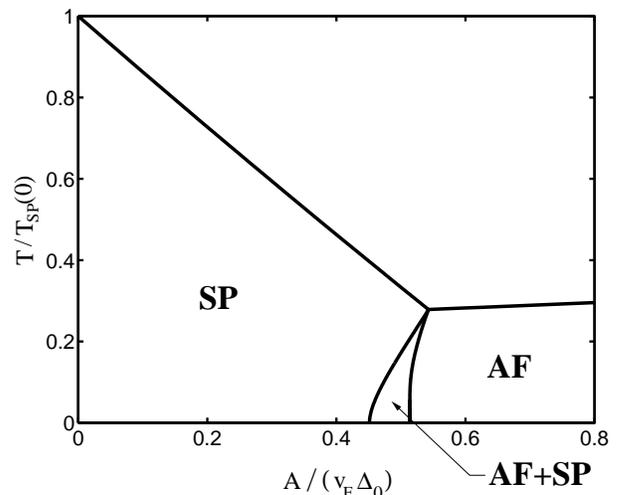}
\caption{
\footnotesize
The phase diagram of the disordered SP system described by
Eqs.(\ref{H_f}),(\ref{scd}), and (\ref{sch}) for
$\lambda_{\Delta} > \lambda_h$.  The dimensionless disorder
strength $A / (v_F \Delta_0)$ is proportional to the
concentration of dopands $x$ (see discussion in the text).  The
temperature is measured in the units of the SP transition
temperature at zero disorder.
\label{phdiag1}}
\end{figure}

From the above we conclude that for $\lambda_h >
\lambda_{\Delta}$ the SP state is less favorable than the AF
state at all values of the disorder strength $A$.  If, on the
other hand, $\lambda_{\Delta} > \lambda_{h}$, a much richer
phase diagram arises, as is observed in Fig.~\ref{phdiag1}.  This
diagram was obtained by numerically solving Eqs.~(\ref{scd}) and
(\ref{sch}) for $\lambda_{\Delta}$ and $\lambda_h$ such that
$T_{N}^0(0) / T_{SP}^0(0) = 1 / 4$, where $T_{SP}^0(A)$
is the SP transition temperature at $h = 0$ and $T_{N}^0(A)$
is the Neel temperature at $\Delta = 0$.  Four phases appear: SP,
AF, mixed SP$+$AF and disordered, separated by second order
transition lines.  At low temperature and weak disorder the
system is in the SP state.  The SP temperature $T_{SP}(A)$
decreases almost linearly with the disorder strength.  In
particular, it can be shown that at small $A$
\begin{equation}
T_{SP}(A) = T_{SP}(0)
\left( 1 - C \frac{A}{v_F \Delta_0} \right),\;\:
C = \frac{\pi^2}{4 \gamma} \approx 1.39,
\label{C1}
\end{equation}
where $\Delta_0$ is the value of $\Delta$ for $T,A = 0$, and
$\gamma = 1.78\ldots$, is the exponential of Euler's constant.

Above some critical disorder strength $A_{N}$, the system
undergoes at $T_{N}(A) < T_{SP}(A)$ a second (Neel)
transition into the mixed state, in which the SP and AF orders
coexist.  This coexistence region becomes narrower when
$\lambda_h$ approaches $\lambda_\Delta$.  $T_{N}$ rapidly
increases with the disorder strength until at some $A = A_{\ast}$
it becomes equal to the SP transition temperature:
$T_{SP}(A_{\ast})=T_{N}(A_{\ast})=T_{\ast}$.  Above
$A_{\ast}$ only AF long-range order exists and the Neel
temperature continues to grow slowly with the disorder strength.

\begin{figure}[htbp]
\centering \leavevmode
\epsfxsize=8cm
\epsfbox{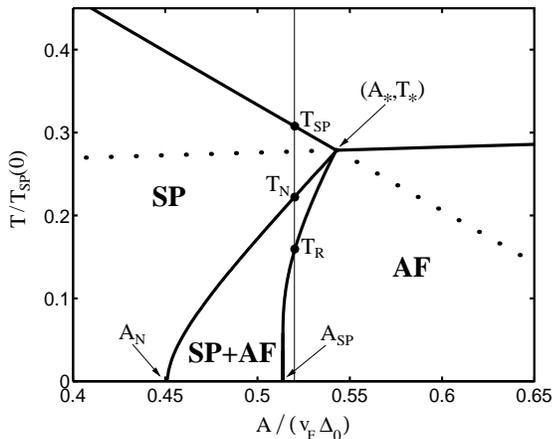}
\caption{
\footnotesize
Detail of the phase diagram Fig.~\ref{phdiag1}.  The vertical
line $A / (v_F \Delta_0) = 0.52$ passes through three
phase-transition points: the SP transition temperature
$T_{SP}$, the Neel temperature $T_{N}$, and the temperature
of re-entrance into the undimerized state $T_{R}$.  Dotted
lines show $T_{SP}^0$ for $A > A_{\ast}$ (the SP transition
temperature calculated at $h=0$) and $T_N^0$ for $A < A_{\ast}$
(the Neel temperature calculated at $\Delta = 0$).
\label{phdiag2}}
\end{figure}

The surprising feature of our phase diagram is the fact
that the disorder strength $A_{SP}$ at which the dimerization
disappears at zero temperature, is smaller than $A_{\ast}$.  This
implies that for $A_{SP}<A<A_{\ast}$ the system experiences
three consecutive transitions as the temperature goes down (see
Fig.\ref{phdiag2}): first the SP transition, next the Neel
transition, and then the ``anti-spin-Peierls'' transition, at
which the SP order disappears.  The re-entrance into the
undimerized state occurs, because the rapid growth of the AF
order parameter below the Neel temperature suppresses the
SP state.

The existence of a re-entrance transition can be further
elucidated by considering the Ginzburg-Landau expansion of
$\Omega = \Omega_f + \frac{\Delta^2}{2 \lambda_{\Delta}}
+ \frac{h^2}{2 \lambda_{h}}$ near the multicritical point
$(T_{\ast},A_{\ast})$:
\begin{eqnarray}
\Omega & = & \alpha_{\Delta}
\left(T - T_{SP}^0(A)\right) \Delta^2 +
\frac{b_{\Delta}}{2} \Delta^4 \nonumber \\
& + & \alpha_{h} \left(T - T_{N}^0(A)\right) h^2 +
\frac{b_{h}}{2} h^4
+ c \Delta^2 h^2 .
\label{OmegaGL}
\end{eqnarray}
In Eq.(\ref{OmegaGL}) the coefficients $\alpha_{\Delta}, \alpha_h
> 0$.  Furthermore, the stability of the system described by
Eq.(\ref{OmegaGL}), requires $b_{\Delta},b_h$, and $D \equiv
b_{\Delta} b_h - c^2$ to be positive.

In the presence of a dimerization (at $A < A_{\ast}$) the Neel
temperature becomes:
\begin{equation}
T_{N}(A) = T_{N}^0(A) - \frac{c}{\alpha_h} \Delta^2 .
\label{T_N(A)}
\end{equation}
As the dimerization suppresses the AF state, $c > 0$.
Similarly, one can find a temperature $T_R(A)$, at which
$\Delta$ becomes zero at nonzero $h$:
\begin{equation}
T_{R}(A) = T_{SP}^0(A) - \frac{c}{\alpha_{\Delta}} h^2 ,
\label{T_R(A)}
\end{equation}
which is the temperature of the re-entrance into the
undimerized state.

To obtain the dependence of $T_{N}(A)$ and $T_{R}(A)$ on
the disorder strength, we find $\Delta$ and $h$ that minimize
$\Omega$ and substitute them into Eqs.(\ref{T_N(A)}) and
(\ref{T_R(A)}). The result is:
\begin{eqnarray}
T_{N}(A) & \approx & T_{\ast} + (A - A_{\ast})
\left[
\frac
{c \alpha_{\Delta} \frac{d T_{SP}^0}{dA} -
b_{\Delta} \alpha_{h} \frac{d T_{N}^0}{dA}}
{c \alpha_{\Delta} - b_{\Delta} \alpha_{h}}
\right]_{A = A_{\ast}},
\nonumber \\ \label{T_NR(A)} \\
T_{R}(A) & \approx & T_{\ast} + (A - A_{\ast})
\left[ \frac
{c \alpha_h \frac{d T_{N}^0}{dA} -
b_h \alpha_{\Delta} \frac{d T_{SP}^0}{dA}}
{c \alpha_h - b_h \alpha_{\Delta}} \right]_{A = A_{\ast}}
\nonumber .
\end{eqnarray}

From Eq.(\ref{T_NR(A)}) and the fact that $T_{N}^0(A)$ increases
with $A$, while $T_{SP}^0(A)$ decreases, it is easy to find that
for
\begin{equation}
c > b_h \frac{\alpha_{\Delta}}{\alpha_h}
\label{ineq}
\end{equation}
both $T_{N}(A)$ and
$T_{R}(A)$ increase linearly with disorder at $A < A_{\ast}$ and
$T_{R}(A) < T_{N}(A)$.  Therefore, inequality (\ref{ineq}) is the
condition for the existence of the re-entrance transition.  In
our model, its validity can be checked analytically for
$\lambda_h \rightarrow \lambda_{\Delta}$, in which case $A_{\ast}
\rightarrow 0$, $\alpha_h \rightarrow \alpha_{\Delta}$, and $c
\rightarrow 2 b_h$.  Our numerical calculations suggest that
condition (\ref{ineq}) is satisfied for all $\lambda_h <
\lambda_{\Delta}$.

Next we compare our phase diagram to experimental data.  At small
dopand concentrations $x$, the observed SP transition temperature
in Cu$_{1-x}$Zn$_x$GeO$_3$ is described by
\begin{equation}
T_{SP}(x) = T_{SP}(0) \left(1 - \alpha x\right) ,
\label{exp}
\end{equation}
where $\alpha \sim 14$ \cite{Hase93.2}.  To compare this to our
result Eq.(\ref{C1}), we have to relate the disorder strength $A$
to the dopand concentration $x$.  This can be done by assuming
that the substitution of Cu by Zn changes the spin exchange
constant by an amount $\sim J_0$.  From Eqs.~(\ref{etaJ}) and
(\ref{Gauss}) we then obtain:
\begin{equation}
A \sim a J_0^2 x .
\end{equation}
Equation (\ref{C1}) then reduces to Eq.(\ref{exp}) with
\begin{equation}
\alpha \sim \frac{J_0}{\Delta_0} C \sim 9 .
\label{alpha}
\end{equation}
Keeping in mind that in our model doping-induced large
fluctuations in some (randomly chosen) spin-exchange constants
are replaced by small fluctuations on all bonds, the agreement
with experiment is good.  Also the even stronger suppression
($\alpha \sim 50$) of the SP phase in Si doped CuGeO$_3$
\cite{Renard,Regnault,Hir,Grenier} may be understood: Si,
substituting Ge, is located between two CuO$_2$-chains and thus
influences two chains simultaneously \cite{KGM}.

One not quite satisfactory aspect of our model is that for
realistic values of parameters the critical doping concentration
$x_c$ at which AF order appears, seems to be too
large as compared with experiment \cite{Grenier}.  This may
be understood from the following considerations: Disorder
enhances the AF susceptibility of spin chains by filling the SP
gap with low-energy spin excitations.  As was shown in
Ref.~\CITE{MK}, with highest probability the excited states
with energy $\varepsilon \ll \Delta$ occur for disorder
fluctuations $\eta(x)$ that have the form of a kink-antikink
pair.  For such a fluctuation, the order parameter $\Delta(x) =
\Delta + \eta(x)$ has reversed sign in a domain of length $R =
(v_F/\Delta) \ln (2 \Delta / \varepsilon)$ between the kink and
the antikink.  The kink and antikink, being fractionally charged
objects \cite{JS}, each carry spin $\frac12$, which together form
a weakly bound singlet.  A low-energy excited state is then
obtained by exciting this singlet into a triplet.  These weakly
bound spins do not contribute to the dimerization, but they can
give rise to AF ordering.  However, for weak disorder (small
$x$), the density of kink-antikink fluctuations is exponentially
small in our model, implying that a critical dopand concentration
$x_c$ is necessary for AF order to appear.  It can be argued that
$x_c$ would be lowered if the model would allow for large
fluctuations on some randomly chosen bonds, in which case the
kink density $\propto x$ \cite{KGM,Zeit}.  Such an extension of
the model, however, makes it much harder to obtain the complete
phase diagram, as it becomes impossible to make the continuum
approximation.

To summarize, we obtained the phase diagram of a disordered SP
system, described by a mean-field model.  We showed that disorder
results in a strong suppression of the SP state and gives rise to
AF long range order, which in a certain range of the disorder
strength coexists with the dimerization.  These results
are in agreement with the experimental data on doped CuGeO$_3$.
Finally, our results indicate the possibility of a
re-entrance transition from the dimerized SP state back into
undimerized state.

This work is supported by the "Stichting voor Fundamenteel
Onderzoek der Materie (FOM)".  We would like to thank Prof.
J.-P.~Renard for providing us with the manuscript of
Ref.~\CITE{Grenier} prior to its publication.


\begin{references}


\bibitem[*]{Perm1} also at G. I. Budker Institute of
Nuclear Physics, 630090 Novosibirsk, Russia.

\bibitem[\dagger]{Perm2} also at P. N. Lebedev Physical
Institute, Leninski prosp. 53, Moscow, Russia.

\bibitem{Hase93.1}
M. Hase, I. Terasaki, and K. Uchinokura, Phys. Rev. Lett.
{\bf 70}, 3651 (1993).

\bibitem{Hase93.2} M. Hase {\em et al.}, Phys. Rev. Lett. {\bf
71}, 4059 (1993).

\bibitem{Renard} J.-P.~Renard {\em et al.}, Europhys. Lett.
{\bf 30}, 475 (1995).

\bibitem{Regnault} L.~P.~Regnault, J.~P.~Renard, G.~Dhalenne, and
A.~Revocolevschi, Europhys. Lett. {\bf 32}, 579 (1995).

\bibitem{Hir} M. Hiroi {\em et al.}, Phys. Rev. B {\bf 55}, R6125
(1997).

\bibitem{Grenier} B. Grenier at al., to be published in Phys.
Rev. B.

\bibitem{Fukuyama} H. Fukuyama, T. Tanimoto, and M. Saito, J.
Phys. Soc. Japan, {\bf 65}, 1182 (1996).

\bibitem{KGM} D. Khomskii, W. Geertsma, and M. Mostovoy, Czech.
Journ. of Phys., {\bf 46}, Suppl. S6, 3229 (1996).

\bibitem{Zeit} M.  Mostovoy and D.  Khomskii,
Z. Physik B {\bf 103}, 209 (1997).

\bibitem{Fabrizio} M. Fabrizio and R. M\'elin, Phys. Rev. Lett.
{\bf 78}, 3382 (1997).
%cond-mat 9701149.

\bibitem{IPS} Y. Imry, P. Pincus, and D. Scalapino, Phys. Rev. B
{\bf 12} 1978 (1975).

\bibitem{Bul} L. N. Bulaevskii, Zh. Eksp. Theor. Fiz. {\bf 43},
968 (1962) [Sov. Phys. JETP {\bf 16}, 685 (1963)].

\bibitem{BBK}
L. N. Bulaevskii, A. I. Buzdin, and D. I. Khomskii, Solid
State Commun. {\bf 27}, 5 (1978).

\bibitem{Dasgupta} C. Dasgupta and S. K. Ma, Phys. Rev. {\bf 22},
1305 (1980).

\bibitem{DFisher} D. Fisher, Phys. Rev. B {\bf 50}, 3799 (1994).

\bibitem{Hyman} R. A. Hyman, K. Yang, R.~N.~Bhatt, and
S.~M.~Girvin, Phys. Rev. Lett. {\bf 76}, 839 (1996).

\bibitem{TLM} H. Takayama, Y. R. Lin-Liu, and K. Maki, Phys. Rev.
{\bf 21}, 2388 (1980).

\bibitem{OE}  A. A. Ovchinnikov and N. S. Erikhman,  Zh. Eksp.
Theor. Fiz. {\bf 73}, 650 (1977) [Sov. Phys. JETP {\bf 46}, 340
(1977)].

\bibitem{Inagaki} S. Inagaki and H. Fukuyama, J. Phys. Soc. Japan
{\bf 52}, 3620 (1983).

\bibitem{Castilla} G. Castilla, S. Chakravarty, and V. J. Emery,
Phys. Rev. Lett. {\bf 75}, 1823 (1995).

\bibitem{MK} M.  V.  Mostovoy and J.  Knoester, Phys.  Lett.  A.
{\bf 235}, 535 (1997).

\bibitem{JS}See, {\em e.g.}, R. Jackiw and J. R. Schrieffer,
Nucl. Phys. B {\bf 190}, 253 (1981).

\end{references}
\end{document}